\documentclass[12pt]{article}
\begin{document}
\title{The Magnetism of Neutron States}
\author{B.G. Sidharth\\
International Institute of Applicable Mathematics \& Information Sciences\\
B.M. Birla Science Centre, Adarsh Nagar, Hyderabad - 500 063, India}
\footnotetext{Based on the invited talk at the NATO Advanced Study Institute, 2003 at KEMER}
\date{}
\maketitle
\begin{abstract}
The recent measurement by Bignami and co-workers of the magnetic field of a neutron star for the first time gives a value that differs by about two orders of magnitude from the expected value. The speculation has been that the nuclear matter in the neutron stars exhibits some exotic behaviour. In this note we argue that this exotic behaviour is an anomalous statistics obeyed by the neutrons, and moreover these considerations lead to a value of the magnetic field that agrees with the observation. The same considerations also correctly give the magnetic fields of the earth and Jupiter.
\end{abstract}
Very recently Bignami and co-workers have been able to observe the magnetic field of an isolated neutron star for the first time$^1$ and get the value $8 \times 10^{10}$ Gauss, which differs by a few orders of magnitude from the expected value$^2$. We would like to draw attention to the fact that infact the value just observed was deduced on the basis of the anomalous statistics obeyed by the neutrons in neutron stars, which behave as a degenerate Fermi gas$^{3,4,5}$.\\
The anomalous statistics arises due to the fact that, as is well known, the energy density $e$ at sub Fermi temperatures is given by
\begin{equation}
e \propto \int^{p_F}_0 \frac{p^2}{2m} d^3 p \propto T^{2.5}_F\label{e1}
\end{equation}
where $p_F$ and $T_F$ are respectively the Fermi momentum and Fermi temperature.\\
Comparing (\ref{e1}) with the well known generalization of the Stefan-Boltzmann Law$^{6,7}$, viz.,
$$e \propto T^{n+1}_F$$
we can see that the assembly of neutrons behaves with the fractal dimensionality $1.5$, leading to the anomalous semionic  statistics.\\
On the other hand if we consider an assembly of $N$ spin half particles, it is known that $N_+$ denoting the number of spin up particles, the magnetization per unit volume is given by
\begin{equation}
M = \frac{\mu (2N_+ - N)}{V}\label{e2}
\end{equation}
where $\mu$ is the electron magnetic moment.\\
The point is that at low temperatures with usual dimensionality, there are as many spin up as spin down particles so that $N_+ \approx \frac{N}{2}$. Equation (\ref{e2}) then shows that the magnetization is negligible. On the other hand if the assembly behaved according to Bose-Einstein statistics, then there would be a large magnetization because all particles would align with their spins, that is effectively $N_+ \approx N$.\\
When we have the anomalous statistics due to the fractional dimensionality, then we would have instead
\begin{equation}
N_+ = \beta N, \frac{1}{2} < \beta < 1,\label{e3}
\end{equation}
Equations (\ref{e2}) and (\ref{e3}) are quite general - they hold for Fermi gases at sub Fermi temperatures. Infact, in a completely different context, for the earth core, they give the correct geomagnetic field$^{4,8}$. In this case the Fermi temperature can be easily calculated to be $10^5$ degrees centigrade, well above the core temperature. Further there are $N \approx 10^{48}$ atoms (mostly iron) in the inner core and we get on using (\ref{e2}) and (\ref{e3}) the correct geomagnetic strength of about $1$ Gauss.\\
Let us consider the magnetic field of Jupiter. Remembering that the core density of Jupiter is of the same order as that of the earth, while the core volume is about $10^4$ times that of the earth, we have $N \sim 10^{52}$, so that the magnetism $MV$, from (\ref{e2}) $\sim 10^4$ times the earth's magnetism, as required.\\
In the case of neutron stars we know that the number density $\frac{N}{V} \sim 10^{31}$ particles per cc$^9$. Moreover the assembly of neutrons is known to be below the Fermi temperature.  Whence using (\ref{e3}) in (\ref{e2}), we get the magnetization $\approx 10^{11}$ Gauss, exactly as observed by Bignami and co-workers, who infact believe that there is indeed exotic behaviour of the matter in neutron stars.\\ \\
{\large {\bf References}}\\ 
1. Bignami, G.,  et al., {\it  Nature}, {\bf 423}, 725, 2003.\\
2. Zeilik, M.,  and  Smith, E., {\it Introductory Astronomy and Astrophysics}, Saunders College Publishing, New York, 1987.\\
3. Sidharth, B.G., {\it J.Stat.Phys.}, {\bf 95} (3/4), 1999, pp.775-784.\\
4. Sidharth, B.G., {\it The Chaotic Universe: From the Planck to the Hubble Scale}, Nova Science Publishers, Inc., New York, 2001.\\
5. Sidharth, B.G., {\it Chaos, Solitons and Fractals}, {\bf 14} (6), October 2002, pp.831-838.\\
6. Reif, F., {\it Fundamentals of Statistical and Thermal Physics}, McGraw Hill, Singapore, 1965.\\
7. Schonhammer, K., and Meden, V., {\it Am.J.Phys.}, {\bf 64} (9), 1996, 1168-1176.\\
8. Sidharth, B.G., {\it Journal of Ind.Geophysics Union}, {\bf Vol.3}, No.2, December 1999, pp.23-24.\\
9. Ohanian, C.H.,  and Ruffini, R., {\it Gravitation and Spacetime}, New York, 1994, p.397.\\ \\

{\large {\bf Appendix}}\\ \\
In 1.5 dimensions we have, with the usual notation, $\mu$ being a multiplicative factor,
\begin{equation}
\frac{N}{V} \approx \mu \int \sqrt{p} dpn_p = \mu \int p'^2 dp' \frac{1}{\frac{1}{z}e^{\beta p'^4} + 1}, p' = \sqrt{p}\label{eA1}
\end{equation}
If we compare (\ref{eA1}) with the usual expression viz.,
\begin{equation}
\frac{N}{V} = \mu \int p^2 dp \frac{1}{\frac{1}{z}e^{\beta p^2}+1}\label{eA2}
\end{equation}
we can see the denominator of the integral in (\ref{eA1}) is much larger that that in (\ref{eA2}), that is the contribution of momentum states in
$$p'^4 \beta > 1$$
is negligible compared to the contribution of momentum states in
$$p'^4 \beta \leq 1$$
Effectively there is a  mono energetic collection of Fermions which has been shown to have anomalous Bosonic behaviour (i.e. the occupied states are comparatively crowded).\\
To appreciate this further, we have for two dimensions from an equation like (\ref{eA1})
\begin{equation}
\mu' z \approx log (1+z)\label{eA3})
\end{equation}
$\mu'$, being again a constant factor. Thus (\ref{eA3}) shows that $z$ is small, corresponding to a low density classical type behaviour, rather than the degenerate Fermi gas Quantum behaviour near Fermi temperatures.
\end{document}